Roles of vortices and turbulent eddies in particle preferential concentration and deposition in the human respiratory tract

Mengtao Li (李梦涛), Yawei Wang (王亚伟), Wentao Feng (冯文韬), Yubo Fan (樊瑜波)

*Beijing Advanced Innovation Centre for Biomedical Engineering, School of Biological Science and Medical Engineering, Beihang University, Beijing, 100083, China*

**Abstract:** Precise targeting of pharmaceutical aerosols is critical to the efficacy and safety of inhaled therapies. This study focused on the influences of vortices and turbulent eddies on preferential concentration and deposition of pharmaceutical aerosols within the human respiratory tract. By reconstructing a high-fidelity respiratory tract geometry spanning the nasal cavity down to sequential bronchi from computed tomography (CT) images and assigning physiologically realistic transient breathing profiles, an unsteady Reynolds-averaged Navier-Stokes (URANS) framework using the shear stress transport (SST) $k$–$\omega$ turbulence model and a stress-blended eddy simulation (SBES) framework of human respiratory tract were built. Beyond flow field analyses, Voronoï diagrams and a set of newly-defined spatial accuracy metrics were adopted to quantitatively characterize particle distributions and localized prediction discrepancies. Spectral analysis revealed that the SBES simulations adequately recovered the prominent Kolmogorov -5/3 inertial scaling law. Such turbulent motion induces pronounced radial turbulent dispersion, forming particle clusters with higher cluster fractal dimensions $D > 1.2$ on the mid-plane of the laryngopharynx. Spatial accuracy metrics uncovered a critical dual deficiency of the URANS model: globally over-smears near-wall deposition by roughly 20%, simultaneously fails to resolve most high-concentration 'hotspots' (the top 5% local extrema). These findings highlighted that the numerical resolution of transient vortices and turbulent eddies constitutes an indispensable prerequisite for accurately predicting of particle transport and deposition in the human respiratory tract.

# I. INTRODUCTION

Inhalation therapy is widely used in the treatment of chronic respiratory diseases[1-3]. The transport and deposition patterns of inhaled particles in the respiratory tract are crucial for the efficacy and safety of the therapy.[4, 5] The human respiratory tract features complex geometries, including high-curvature bends and asymmetrical continuous bifurcations. Combined with the periodic oscillatory airflow during breathing, these factors induce flow separation and periodic laminar-to-turbulent transitions, thereby complicating the transport and deposition of inhaled drugs[6-9]. Studies suggest that non-uniform particle deposition driven by complex flows and particle inertia may play important roles in lung cancer development. Balásházy et al. utilized the maximum enhancement factor to investigate local particle deposition patterns in the bronchi, finding that high-concentration deposition at bronchial bifurcations may correlate with lung cancer[10]. Researchers have conducted extensive experimental and numerical studies on respiratory particle transport and deposition, with computational fluid dynamics (CFD) playing a crucial role[11].

Reynolds-averaged Navier-Stokes (RANS) methods remain the dominant numerical framework for modeling respiratory fluid dynamics [6, 7, 11-14]. In recent years, numerical simulations have gradually shifted toward high-fidelity anatomical models of localized respiratory segments, evolving from steady-state computations to unsteady Reynolds-averaged Navier-Stokes (URANS) frameworks.[15, 16]. The URANS models have the ability to incorporate complex transient breathing patterns, ranging from simple sinusoidal waveforms and actual physiological flow profiles to specialized respiratory maneuvers like sniffing or breath-holding. Kuga et al. contrasted particle deposition behaviors under steady and unsteady flow conditions, demonstrating that periodic unsteady simulations yield more accurate predictions of respiratory particle deposition.[11] Imai et al. and Calmet et al. investigated the effects of specific maneuvers, such as breath-holding and sniffing, on particle transport and deposition in the human respiratory tract[14, 17-20]. With a full-airway URANS numerical model paired with an in vitro experimental platform, our team quantified the deviations between URANS predictions and measured in vitro data.[21] The outcomes indicated that although the particle transport fractions computed by URANS across 17 tertiary bronchi were broadly consistent with the mean values from in vitro tests, the experimental data exhibited substantial variability among individual bronchi. These results implied that such discrepancies might originate from interactions between multiscale eddies and particles — physical mechanisms that URANS is intrinsically incapable of resolving.[22]

Classical experiments and numerical simulations of channel and pipe flows have revealed that turbulent eddies would induce local particle accumulation (hotspots)[22-28]. Such particle accumulations exhibit topological morphologies associated with vortex structures, suggesting that local particle accumulation in turbulence is driven by multi-scale vortical structures[29-33]. To investigate the preferential concentration of particles within homogeneous isotropic turbulence in a channel flow, researchers usually adopted the Voronoï diagrams to quantify how the microscale Reynolds number, Stokes number, and particle volume fraction influence the generation of particle clusters and void structures, alongside their corresponding statistical properties[23, 33]. Recently, Cui et al. established a large eddy simulation (LES) model and employed the dynamic mode decomposition (DMD) method to study the dynamic evolution of vortices and turbulent eddies in different physiological regions of the human respiratory tract[34-37]. They found the evolution of vortices is governed by anatomical structural features and breathing frequency, and these vortices exert important influence in the respiratory tract from the glottis down to the bronchi without considering particle transportation and deposition. Wedel compared the differences in simulated flow field evolution and particle deposition between URANS model and detached-eddy simulation (DES) model in the human respiratory tract, while influences of vortices and turbulent eddies in the transport and distribution of drug particles were not studied[38].

This study investigated the mechanisms by which transient vortices and turbulent eddies affect the preferential concentration and deposition dynamics of drug particles in the human respiratory tract. By constructing an anatomical geometric model extending from the nasal cavity to the continuous bronchi and applying physiological breathing conditions, we systematically compared the differences in the results of flow fields and particle dynamics between the URANS model and the stress-blended eddy simulation (SBES) model[39]. Moreover, this study employed the Voronoï diagrams and a set of newly-defined spatial accuracy metrics to characterize particle distribution. This study provided some insights on how the fidelity of turbulence resolution governs radial turbulent diffusion, particle trajectories, and the topological arrangement of high-concentration deposition hotspots.

# II. METHODS

## A. Geometric reconstruction of the holistic airway

Figure 1 illustrates the research setup, computational geometry, and physiological and physical conditions adopted in this study. A high-fidelity anatomical airway model extending from the bilateral anterior nares to the third-generation bronchi was reconstructed from computed tomography (CT) images of a healthy adult male acquired at a matrix size of 512×512 over 417 slices. Image segmentation and geometric reconstruction were performed using 3D Slicer (version 5.8.1). The computational domain comprised the nasal cavity, nasopharynx, oropharynx, larynx, trachea, and tracheobronchial tree, terminating at 17 bronchial outlets. The reconstructed geometry and representative local mesh configurations are shown in Fig. 1b. A physiologically realistic total volumetric flow-rate waveform was obtained by Aeonmed Co., Ltd. using an ASL 5000 active lung simulator. After conversion to the corresponding mass-flow rate, the waveform was prescribed across the bilateral anterior nares such that the combined inlet flow matched the measured total flow rate (Fig. 1c). A pressure-outlet condition was applied independently at each terminal bronchus, while a no-slip condition was imposed on all airway walls. The size distribution of the injected aerosol particles is presented in Fig. 1d.

## B. Numerical simulation framework

### *1. Transient aerodynamic modeling*

To satisfy the different spatial-resolution requirements of the two turbulence-modeling approaches, two sets of unstructured tetrahedral meshes were generated. The SBES simulation employed a high-resolution mesh containing approximately 13 million elements, whereas the URANS simulation using the shear stress transport (SST) $k$–$\omega$ turbulence model employed a mesh containing approximately 3 million elements. Eight layers of prismatic boundary-layer meshes were generated along all airway walls to resolve near-wall velocity gradients, with additional local refinement applied near the bronchial carinas. Mesh convergence was assessed separately for both modeling approaches. For the SBES simulation, the mesh was further evaluated in terms of its turbulence-resolving capability, including its ability to recover the inertial subrange in the turbulent energy spectrum[40-42].

Discrete sampling locations and analysis regions were defined within the respiratory geometry to characterize flow evolution and particle clustering, as illustrated in Fig. 2. The upper-airway configuration is shown in Fig. 2a, where the laryngeal constriction accelerates the flow and promotes downstream flow separation. Three monitoring probes, P1–P3, and four regions of interest, A1–A4, were therefore positioned along the laryngopharyngeal midplane shown in Fig. 2b to quantify the transient flow dynamics and particle distributions. Further downstream, airflow through the tracheobronchial bifurcations generates secondary vortical motions and enhances inertial particle impaction. To examine these mechanisms, three longitudinal midplanes, BP0, LBP1, and RBP1, were selected to characterize flow division through the bifurcations. In addition, three transverse cross sections, CS1–CS3, were defined to analyze secondary-flow structures and near-wall flow development, as shown in Fig. 2c.

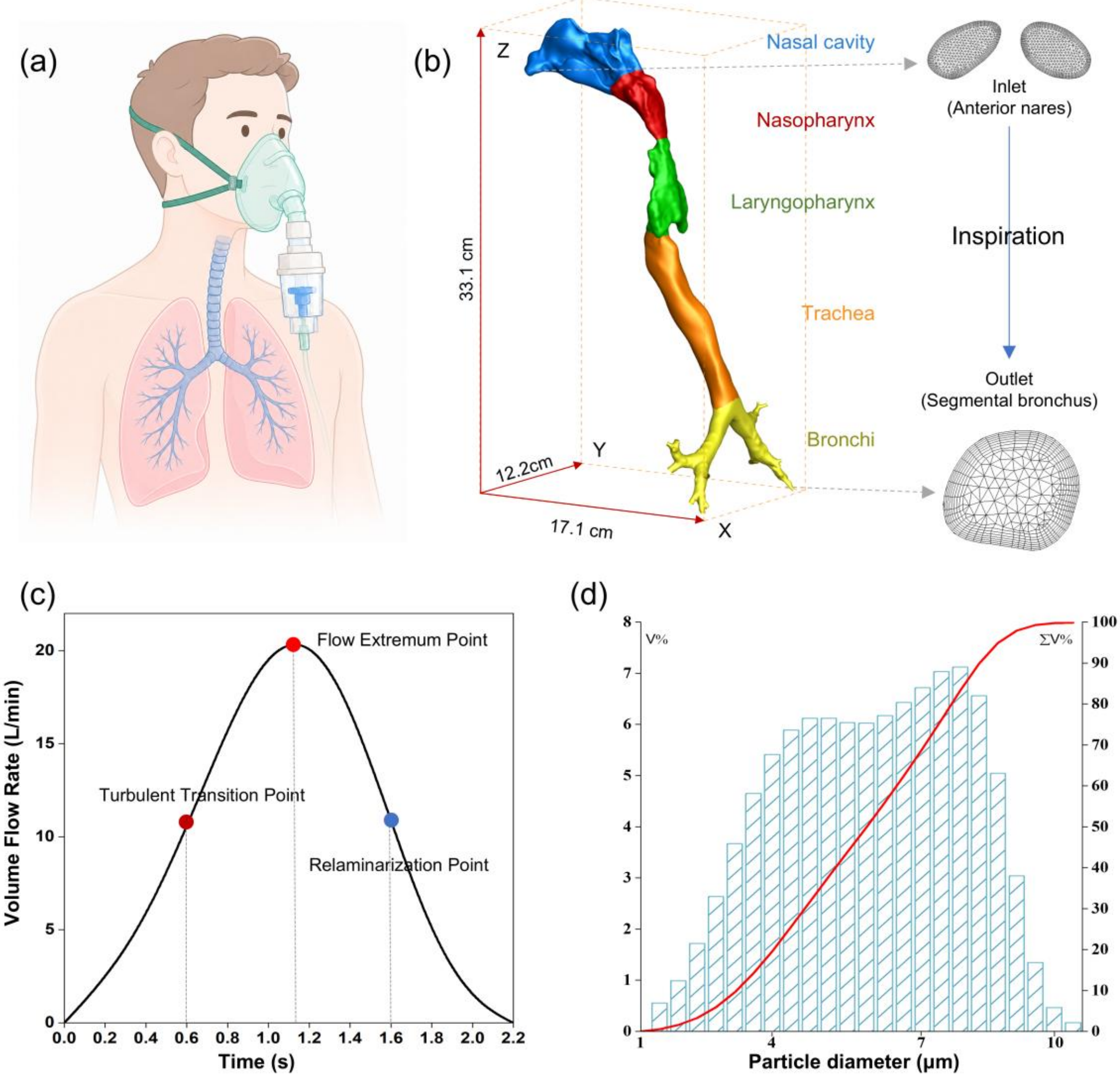

FIG. 1. Research setup of this study. (a) schematic illustration of inhalational therapy; (b) reconstructed high-fidelity geometry of the human respiratory tract; (c) physiologically realistic mass flow rate profile driving the inspiratory phase; (d) droplet size distribution fitted from a commercial nebulizer.

## *2. Numerical setup and discretization*

To systematically evaluate transient flow evolution and particle deposition characteristics, the unsteady Reynolds-averaged Navier-Stokes (URANS) framework using the shear stress transport (SST) $k$–$\omega$ turbulence model and a stress-blended eddy simulation (SBES) framework were deployed in this study. Functioning as a hybrid scale-resolving methodology, SBES utilizes a dynamic blending function to enforce the SST $k$–$\omega$ closure within attached near-wall boundary layers, thereby preserving numerical stability. Within the separated bulk flow, it seamlessly transitions to the wall-adapting local eddy-viscosity (WALE) subgrid-scale (SGS) model to explicitly resolve energy-containing transient eddies. The incompressible continuous phase is governed by the following mass and momentum conservation equations:

$$\frac{\partial}{\partial x_i}(\bar{u}_i) = 0, \tag{1}$$

$$\frac{\partial}{\partial t}(\bar{u}_i) + \frac{\partial}{\partial x_j}(\bar{u}_i\bar{u}_j) = -\frac{1}{\rho}\frac{\partial \bar{p}}{\partial x_i} + \frac{\partial}{\partial x_j}\left[\nu\left(\frac{\partial \bar{u}_i}{\partial x_j} + \frac{\partial \bar{u}_j}{\partial x_i}\right)\right] - \frac{\partial \tau_{ij}}{\partial x_j} \tag{2}$$

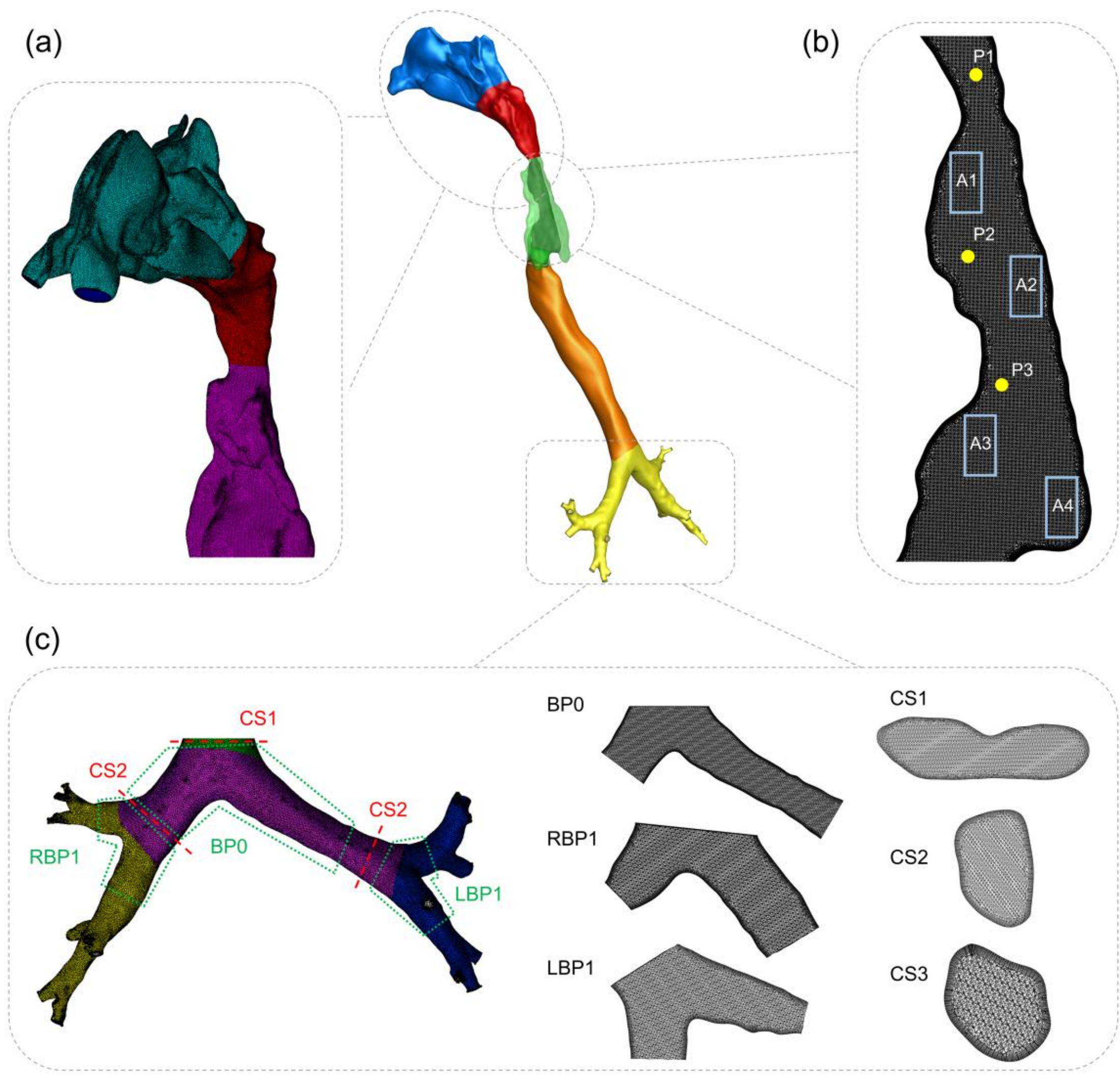

FIG. 2. Global morphology and targeted spatial observation configurations of the respiratory tract. (a) geometrical partition and exterior morphology of the upper respiratory tract; (b) localized laryngopharyngeal mid-plane featuring discrete monitoring probes (P1–P3) and regional observation areas (A1–A4) for voronoï evaluation; (c) hierarchical observation strategy within the lower tracheobronchial tree, detailing the spatial locations and local grid topologies of the longitudinal bifurcation mid-planes (BP0, LBP1, RBP1) and transverse cross-sections (CS1, CS2, CS3).

$u_i$, $\rho$ and $p$ denote the flow velocity vector, fluid density, and static pressure, respectively, while $\tau_{ij}$ represents the unresolved Reynolds stress tensor; the overbar in scalar $\varphi$ signifies the Reynolds-averaging operation within the RANS framework, whereas it denotes the spatial filtering operation within the LES paradigm. To achieve closure for the unknown stress tensor $\tau_{ij}$ in the momentum equations, the Boussinesq eddy-viscosity hypothesis is invoked, strictly coupling the turbulent stress with the resolved mean velocity gradient:

$$\tau_{ij} = -2v_t\bar{S}_{ij} + \frac{2}{3}k\delta_{ij} \tag{3}$$

where $\bar{S}_{ij}$ designates the resolved mean strain rate tensor, and $\delta_{ij}$ is the Kronecker delta.

SBES blends the near-wall RANS stress and the subgrid-scale stress through the shielding function

$$\tau_{ij}^{SBES} = f_b\tau_{ij}^{RANS} + (1 - f_b)\tau_{ij}^{LES} \tag{4}$$

$0 \leqslant f_b \leqslant 1$. This shielding mechanism fundamentally prevents unphysical grid-induced separation (GIS), ensuring a stable and physically consistent RANS-to-LES transition. Within the resolved LES region, the WALE formulation is utilized for the subgrid-scale (SGS) closure. Compared to the classical Smagorinsky model, WALE intrinsically recovers proper near-wall asymptotic behavior without introducing excessive artificial dissipation.

The pressure–velocity coupling was handled using the pressure-implicit with splitting of operators (PISO) algorithm. Cell gradients were reconstructed using the least-squares cell-based method, while pressure interpolation was performed using the PREssure STaggering Option (PRESTO!) scheme. For the SBES simulation, the convective terms in the momentum equations were discretized using the bounded central-differencing scheme to reduce numerical dissipation while maintaining boundedness. For the URANS simulation, a second-order upwind scheme was used for the momentum equations. The turbulence-transport equations were discretized using a second-order upwind scheme. Temporal derivatives were discretized using a second-order implicit scheme. Both simulations were initialized from a quiescent flow field and advanced using a uniform time step of $4.0\times10^{-5}$ s, for which the maximum Courant–Friedrichs–Lewy number remained below 1.0 throughout the simulations. Within each time step, iterations were continued until the scaled residuals of all monitored governing equations fell below $10^{-3}$ and the global mass imbalance satisfied the prescribed conservation criterion.

*3. Lagrangian discrete phase modeling*

Particle transport was simulated within an Eulerian–Lagrangian framework using the discrete phase model (DPM). The dispersed phase remained dilute, with a global particle volume fraction below 1%, and particle–particle interactions were therefore neglected. Owing to the low particle loading, feedback from the particles to the carrier flow was also neglected, and one-way coupling was adopted. The injected particle-size distribution shown in Fig. 1d was represented using a Rosin–Rammler distribution.

Turbulent particle dispersion was treated according to the turbulence-modeling framework. In the URANS simulation, the discrete random walk (DRW) model was used to represent the effects of unresolved turbulent velocity fluctuations on particle trajectories. The DRW model was not applied in the SBES simulation, where particle dispersion was driven by the resolved unsteady velocity field in the scale-resolving regions. Introducing an additional stochastic DRW contribution in these regions could double-count the effects of resolved turbulence and artificially enhance particle dispersion. The contribution of unresolved subgrid-scale fluctuations to particle dispersion was therefore neglected in the SBES calculation.

## C. Data analysis

Before quantitative analysis, the Lagrangian particle data were mapped onto the corresponding analysis surfaces. For the airborne-particle analysis, instantaneous particle positions within a prescribed sampling slab centered on each selected analysis plane were projected onto that plane to characterize preferential concentration. For the deposition analysis, particle deposition locations accumulated over the sampling interval were mapped onto the airway walls to characterize the spatial distribution of deposited particles. Voronoï tessellation was applied to the projected airborne-particle coordinates, and the inverse Voronoï-cell area was used as a local measure of the projected particle number density.

Each Voronoï-cell area $A_i$ was normalized by the mean cell area $\bar{A}$ as $v = A_i / \bar{A}$. The relative probability density function (PDF) is defined as the ratio between the measured cell-area PDF and its corresponding Poisson–Voronoï reference distribution. Accordingly, values greater than unity signify a surplus of cells at the given size when compared with a spatially random particle configuration. A lower physical cutoff, $v_\eta = \eta^2 / \bar{A}$ was applied before determining the classification threshold, where $\eta$ was the Kolmogorov length scale. The first downward crossing of the relative PDF through unity defined the particle-rich-cell threshold, whereas the last upward crossing defined the particle-poor-cell threshold. Cells between these two thresholds were classified as intermediate cells.

To quantify the prediction biases between URANS and SBES, a set of SBES-referenced asymmetric spatial accuracy metrics was defined. Two threshold-based deposition masks were evaluated. The global mask represented the effective deposition region identified using a prescribed baseline threshold, whereas the hotspot mask represented locations where the local deposition intensity exceeded the 95th percentile of the SBES reference field. For each comparison, the same threshold was applied to both the SBES and URANS fields. Relative to the SBES reference mask, regions not identified by URANS were classified as omissions or false negatives, whereas regions identified only by URANS were classified as spurious predictions or false positives.

Four metrics were defined to quantify these prediction biases in terms of deposition area coverage and mass loading. The missed-area and missed-mass fractions quantify the proportions of the SBES reference region and reference mass not reproduced by URANS. Conversely, the false-area and false-mass fractions quantify the proportions of the URANS-predicted region and predicted mass that do not overlap the corresponding SBES reference. The same four metrics were applied at both the global-deposition and hotspot levels, thereby distinguishing broad differences in deposition coverage from localized biases in high-concentration regions.

$F_{\mathrm{A,Missed}}$ was defined to quantify the fraction of the SBES reference deposition area that is not reproduced by URANS:

$$F_{\mathrm{A,Missed}} = \frac{\mathrm{A_{SBES}} \setminus \mathrm{A_{URANS}}}{\mathrm{A_{SBES_Total}}} \tag{5}$$

$F_{\mathrm{M,Missed}}$ was defined to quantify the fraction of the SBES reference deposited mass located within regions omitted by URANS:

$$F_{\mathrm{M,Missed}} = \frac{\sum \mathrm{M_{SBES}} \in (\mathrm{A_{SBES}} \setminus \mathrm{A_{URANS}})}{\mathrm{M_{SBES_Total}}} \tag{6}$$

$F_{\mathrm{A,False}}$ was defined to quantify the fraction of the URANS-predicted deposition area that does not overlap the SBES reference region:

$$F_{\mathrm{A,False}} = \frac{\mathrm{A}_{\mathrm{URANS}} \setminus \mathrm{A}_{\mathrm{SBES}}}{\mathrm{A}_{\mathrm{URANS_Total}}} \tag{7}$$

$F_{\mathrm{M,False}}$ was defined to quantify the fraction of the URANS-predicted deposited mass located within regions not supported by the SBES reference:

$$F_{\mathrm{M,False}} = \frac{\sum \mathrm{M}_{\mathrm{RANS}} \in (\mathrm{A}_{\mathrm{URANS}} \setminus \mathrm{A}_{\mathrm{SBES}})}{\mathrm{M}_{\mathrm{URANS_Total}}} \tag{8}$$

# III. RESULTS

## A. Turbulent characteristics of the laryngeal jet

Figure 3 compared the instantaneous velocity-magnitude fields on the laryngopharyngeal midplane during the acceleration phase at t = 0.6s, peak inhalation at t = 1.1s, and the deceleration phase at t =1.6s. The URANS results retain a relatively smooth laryngeal jet with a continuous high-velocity core and limited spatial variability in the downstream wake, reflecting the Reynolds-averaged treatment of turbulent fluctuations. In contrast, the SBES fields exhibit pronounced shear-layer instability, lateral deformation of the jet, and downstream breakup into transient vortical structures and finer-scale turbulent motions. These differences become particularly evident near peak inhalation and during deceleration. The unsteady structures resolved by SBES enhance cross-stream momentum exchange and provide the hydrodynamic basis for the radial turbulent diffusion and particle preferential concentration examined in the following sections.

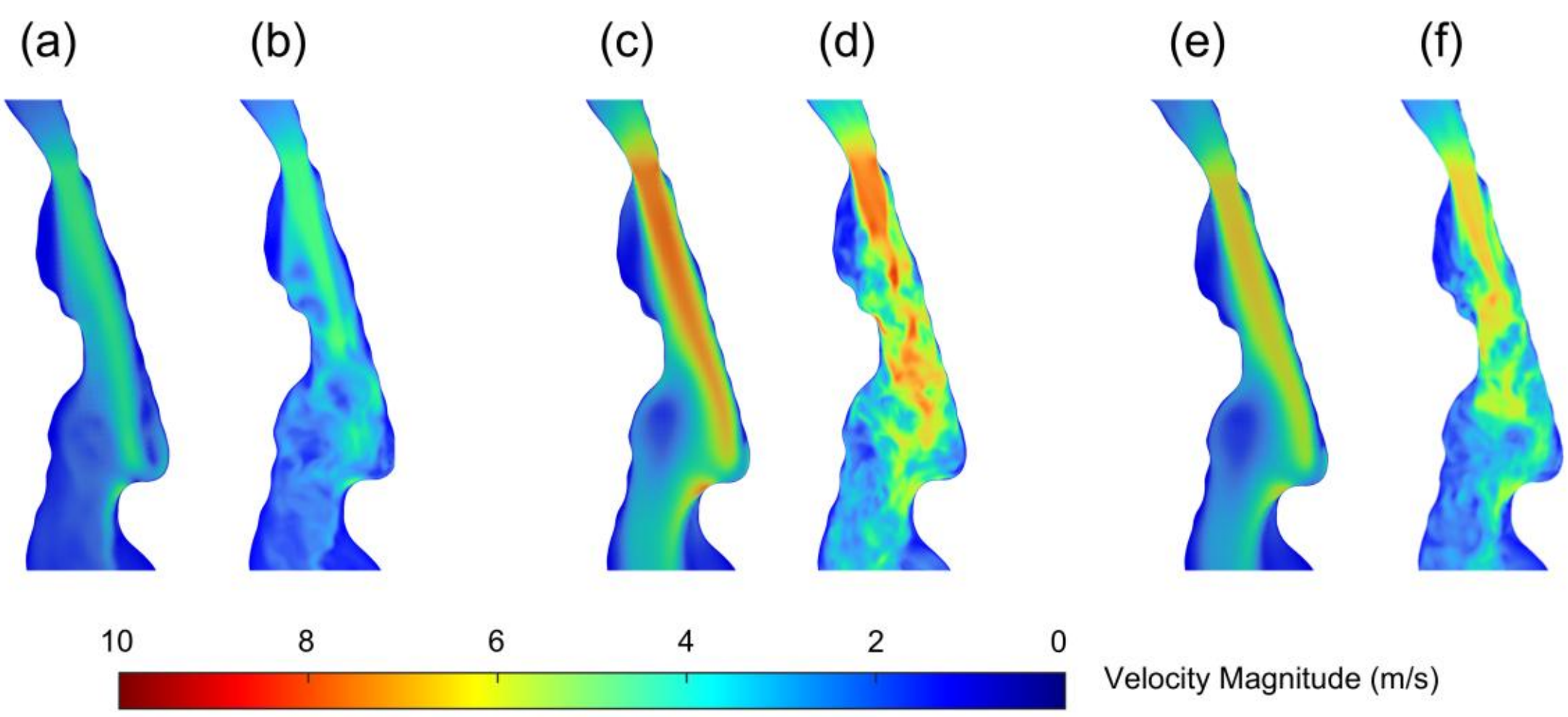

FIG. 3. Instantaneous velocity-magnitude fields on the laryngopharyngeal midplane predicted by URANS in panels (a), (c), and (e) and by SBES in panels (b), (d), and (f), at $t$ = 0.6 s, 1.1 s, and 1.6 s, respectively.

Figure 4 compares the time histories of the monitored velocity component at P1–P3, with the SBES and URANS results presented in Fig. 4a–c and Fig. 4d–f, respectively. The SBES velocity signals exhibit sustained rapid fluctuations during most of the inhalation stage, with the fluctuation amplitude increasing markedly from P1 to P3. This downstream amplification indicates the progressive development of unsteady motions within the laryngeal jet wake. In contrast, the URANS velocity signals retain the large-scale temporal variation of the flow but are substantially smoother, although intermittent transient excursions remain at P2 and P3. This difference arises from the distinct turbulence treatments of the two approaches: SBES explicitly resolves energy-containing transient eddies in the scale-resolving regions, whereas URANS represents most turbulent velocity fluctuations through the Reynolds-averaged turbulence closure. The growth of the SBES fluctuations downstream is consistent with the shear-layer deformation and wake fragmentation observed in Fig. 3 and is further quantified through the spectral analysis in Fig. 5.

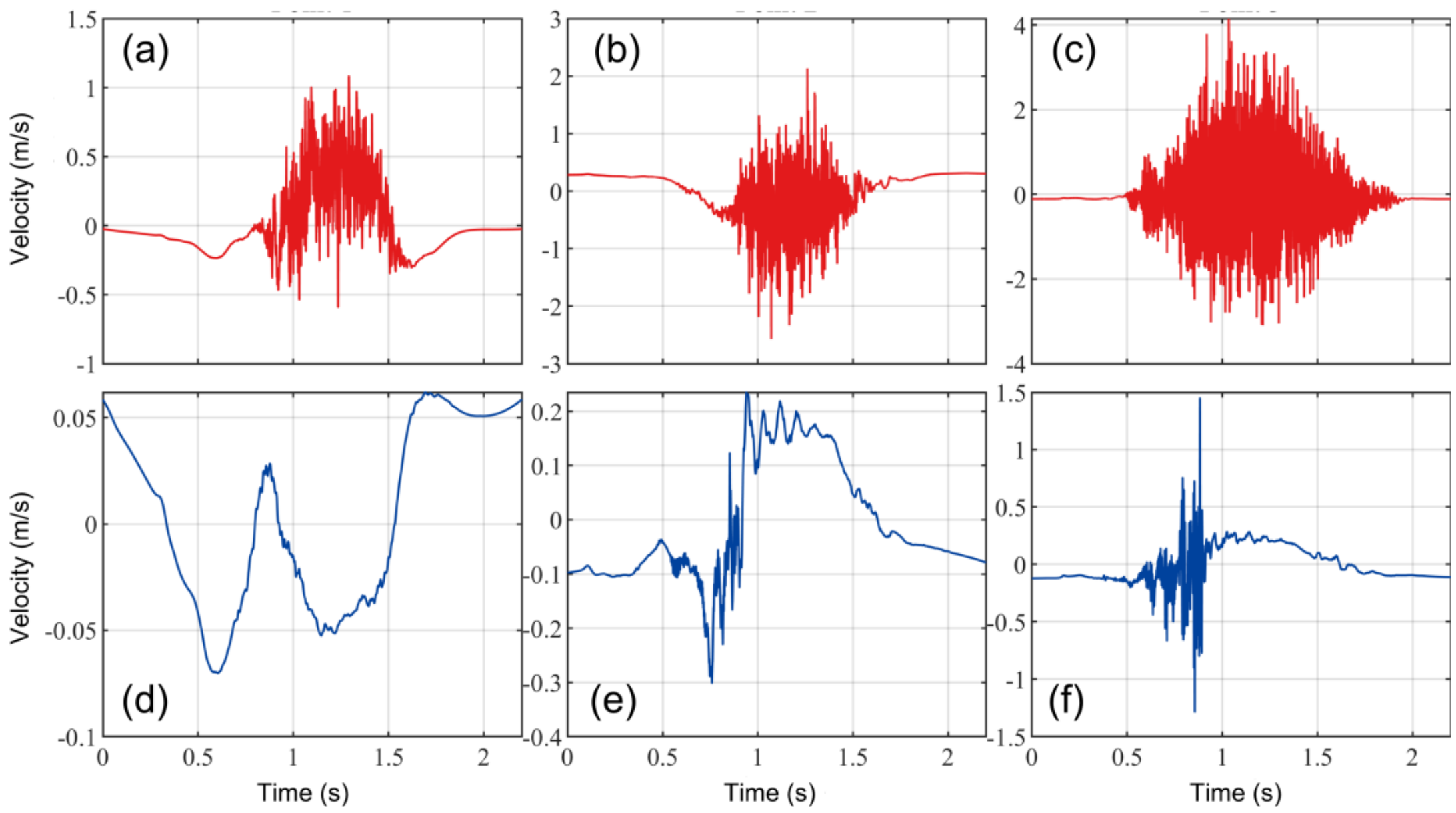

FIG. 4. Time-domain evolution of velocity fluctuations at the three designated monitor points. The transient profiles for points 1, 2, and 3 are detailed in (a, d), (b, e), and (c, f), respectively. Red and blue lines denote the SBES and URANS predictions, respectively.

Figure 5 compared the temporal and spectral characteristics of the monitored velocity component at P1–P3 along the laryngeal jet over physical time of 0.6 s ~1.6 s. The results for P1, P2, and P3 are grouped in Fig. 5a–c, Fig. 5d–f, and Fig. 5g–i, respectively. At P1, the SBES signal exhibits intermittent velocity fluctuations of relatively limited amplitude, while its spectral content remains concentrated primarily at lower frequencies associated with large-scale energy-containing motions. Correspondingly, the power spectral density (PSD) did not exhibit a sustained frequency range parallel to the Kolmogorov -5/3 reference slope. Further downstream at P2 and P3, the SBES velocity signals exhibited stronger and more persistent turbulent velocity fluctuations. Their Fourier amplitude spectra extend over a substantially broader frequency range, indicating the progressive development of smaller temporal scales as the laryngeal jet wake evolves downstream. Portions of the corresponding PSDs exhibit an approximately $f^{-5/3}$ scaling over a finite frequency range, consistent with the emergence of an inertial subrange and the transfer of turbulent energy toward higher frequencies. In contrast, the URANS velocity signals remain substantially smoother, and their Fourier amplitudes and PSDs decay rapidly throughout the intermediate- and high-frequency ranges. These spectral differences demonstrate that SBES resolves a broader range of turbulent eddies, whereas URANS predominantly retains the larger-scale, lower-frequency evolution of the Reynolds-averaged flow.

## B. Jet-driven preferential concentration in the laryngopharynx

The unsteady laryngeal jet governs the instantaneous spatial organization of particles within the laryngopharynx. Fig. 6 compared the particle distributions predicted by SBES and URANS at peak inhalation, i.e., t = 1.1s. The SBES result exhibited pronounced preferential concentration, with particle-rich clusters separated by distinct particle-poor regions. This spatial intermittency is induced by the transient vortices and turbulent eddies resolved in the laryngeal jet. In contrast, the particle distribution predicted by the URANS model is substantially more homogeneous, with weaker contrasts between clustered and depleted regions. Although the DRW model introduces stochastic turbulent dispersion into the URANS model's particle trajectories, it did not reproduce the spatially correlated particle–eddy interactions resolved in SBES.

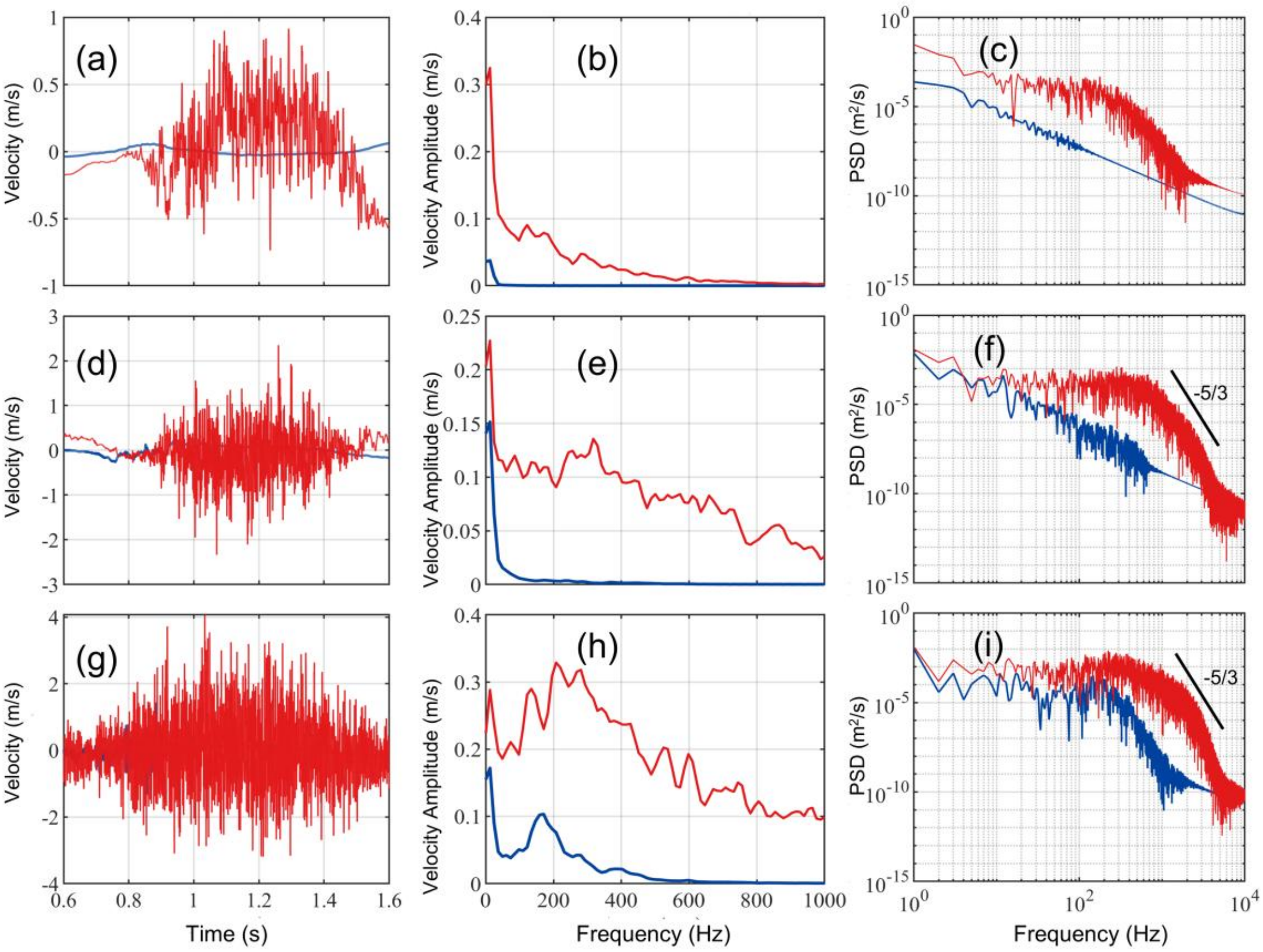

FIG. 5. Temporal and spectral analyses of the monitored velocity component at P1–P3. Results for monitoring points P1, P2, and P3 are grouped in (a–c), (d–f), and (g–i), respectively, with each group detailing the time-domain (a, d, g), frequency-domain (b, e, h), and power spectral density (PSD) (c, f, i) characteristics. Red and blue lines denote the SBES and URANS predictions, respectively.

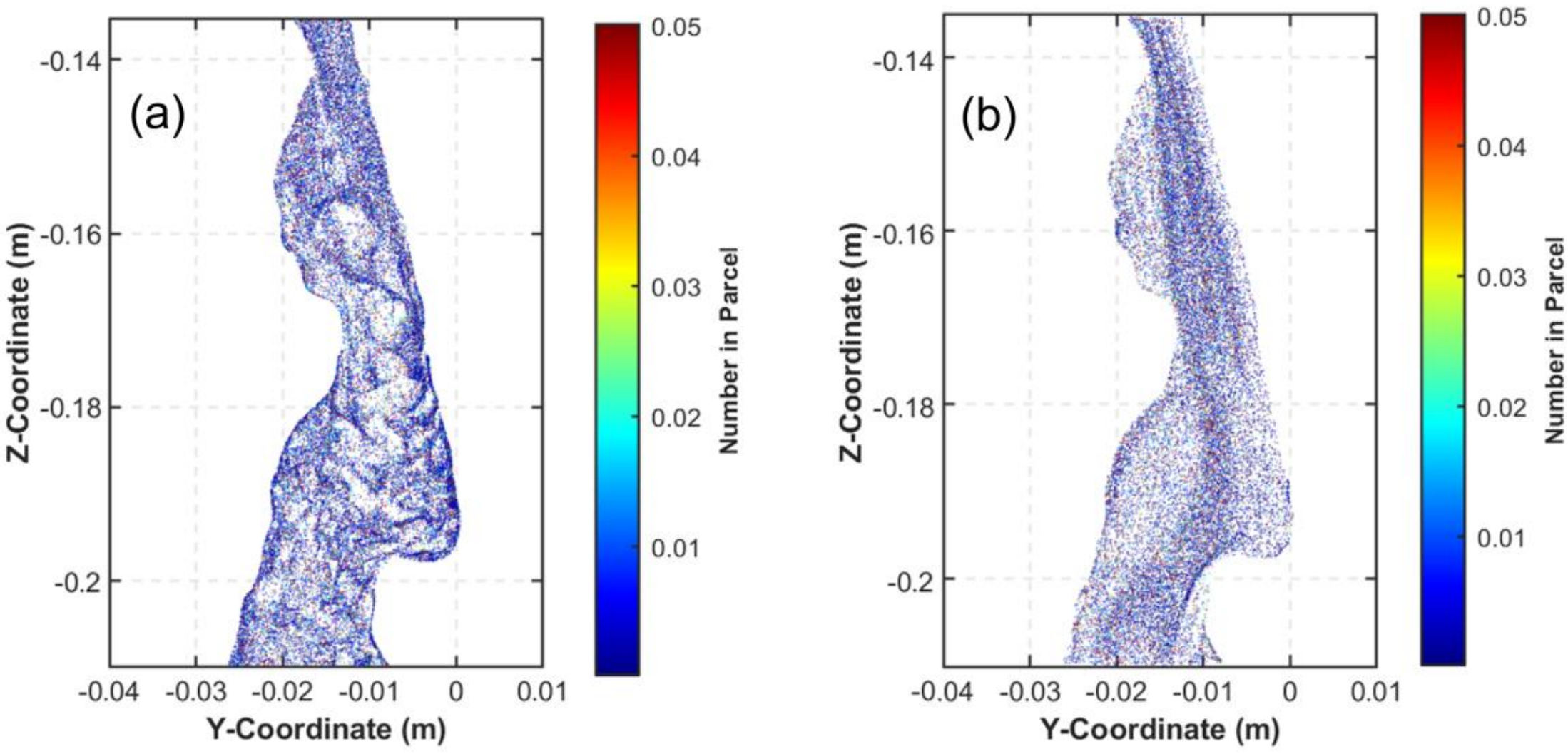

FIG. 6. Instantaneous particle distributions on the laryngopharyngeal midplane at peak inhalation (t = 1.1 s) predicted by (a) the SBES model and (b) the URANS model.

To quantify the preferential concentration phenomenon observed in Fig. 6, Voronoï analysis was applied to the four regions of interest, A1–A4, defined in Fig.2b. Fig. 7 compares the URANS and SBES predictions using three complementary descriptors: Voronoï-cell morphology, the relative PDF of the normalized area $V$ and the perimeter–area scaling of the identified particle clusters. The relative PDF was normalized by the corresponding Poisson reference distribution.

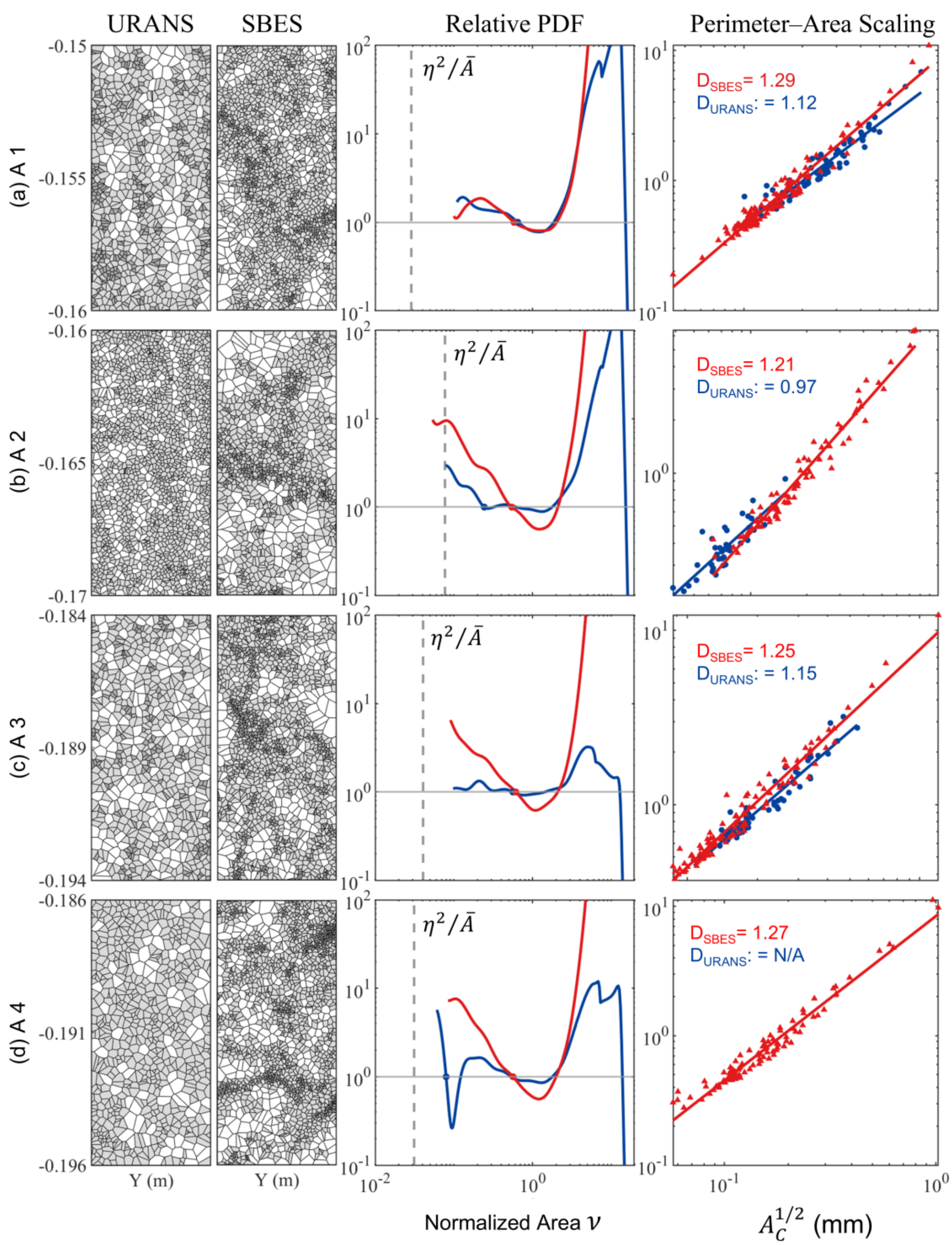

FIG. 7. Voronoï characterization of instantaneous particle distributions in four representative regions: (a) A1, (b) A2, (c) A3, and (d) A4. In each row, the columns show the URANS and SBES tessellations, the relative PDFs of normalized Voronoï-cell area, and the perimeter–area scaling of connected particle clusters. Black, gray, and white cells denote particle-rich, intermediate, and particle-poor regions, respectively; red and blue denote SBES and URANS. In the relative-PDF panels, the solid gray line marks unity and the dashed line marks the normalized cutoff. N/A indicates that the prescribed minimum cluster-number and scale-range criteria for establishing the perimeter–area scaling relation were not satisfied.

As depicted in Fig.7, the SBES tessellations generally exhibit a broader coexistence of small and large Voronoï cells, with the differences becoming particularly pronounced in A2–A4. Small cells correspond to locally particle-rich regions, whereas large cells represent particle-poor regions. Consistent with these morphological features, the SBES relative PDFs exhibit a stronger excess in the small-cell range and extend over a broader large-cell range. By comparison, the URANS distributions show weaker departures from the random reference and generally terminate earlier at large normalized areas. These results indicate that URANS underpredicts both the preferential clustering of particles and the occurrence of extended particle-poor regions.

It was also depicted in Fig.7 that, the fractal analysis further distinguishes the geometries of the connected particle clusters predicted by the two approaches. The cluster fractal dimension $D$ was obtained from the perimeter–area scaling relation $P_c \propto A_c^{D/2}$, corresponding to the slope of $\log P_c$against $\log A_c^{1/2}$. The SBES clusters yield fractal dimensions of 1.21-1.29 across A1–A4, whereas the URANS values range from 0.97 to 1.15 in A1–A3. Because a compact cluster with a relatively smooth boundary gives $D$ close to unity, the consistently larger SBES values indicate more irregular cluster boundaries and greater multiscale geometric complexity. In A4, the URANS distribution exhibits only a limited excess of small Voronoï cells and does not produce a cluster population, and thus does not satisfy the prescribed minimum cluster-number and scale-range criteria to establish a fractal scaling relation. The Voronoï morphology, relative PDFs, and fractal scaling in Fig.7 demonstrated that the SBES model predicts stronger spatial intermittency and more complex particle-cluster morphologies than the URANS model, which were in consistent with the particle–eddy interactions generated by the resolved transient vortices and turbulent eddies.

## C. Particle distributions at bronchial bifurcations

Fig.8 compared the instantaneous velocity-magnitude fields and particle distributions on the longitudinal bifurcation midplanes RBP1, BP0, and LBP1. Although the two models predict broadly similar large-scale velocity distributions through the bifurcations, their particle patterns differ markedly. The URANS results exhibit persistent particle-poor zones within the airway core, together with elongated particle-rich streaks along preferred near-wall pathways, particularly on BP0 and LBP1. In contrast, the SBES predictions show reduced central depletion and less continuous near-wall streaking, with a larger fraction of particles redistributed toward the airway core.

These differences arise from the distinct treatments of turbulent particle transport in the two approaches. In the URANS calculation, particle trajectories are governed by the Reynolds-averaged carrier flow together with stochastic velocity fluctuations introduced through the DRW model. Although the DRW treatment represents unresolved turbulent dispersion statistically, it does not reproduce the spatially correlated particle–eddy interactions associated with resolved transient vortices. Consequently, the spatial confinement imposed by the time-averaged secondary flows and inertial migration remains persistent, directing particles along preferred near-wall trajectories. In the SBES simulation results, the resolved transient vortices and turbulent eddies generate radial turbulent diffusion that intermittently disrupts these trajectories and transports particles back toward the airway core. The resulting particle distributions therefore exhibit weaker large-scale segregation and more pronounced spatial intermittency, consistent with the broader range of turbulent velocity fluctuations identified in Fig. 4 and Fig. 5.

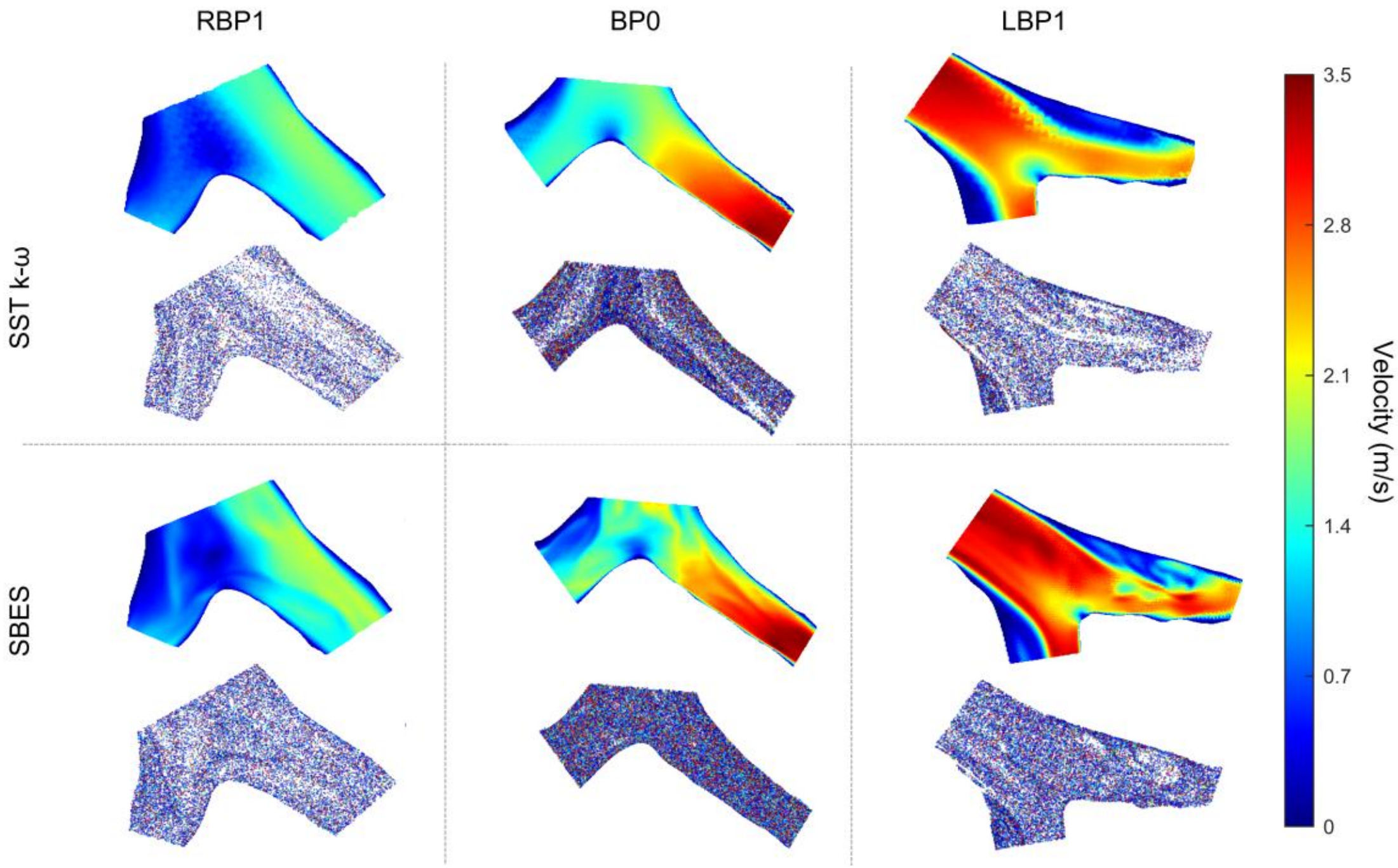

FIG. 8. Instantaneous velocity-magnitude fields and particle distributions on the longitudinal bifurcation midplanes RBP1, BP0, and LBP1. The upper and lower rows show the URANS and SBES results, respectively.

Fig.9 compares the instantaneous velocity-magnitude fields and particle distributions on the transverse sections CS1–CS3. The differences between the two approaches are most evident at CS1 and CS2. At CS2, the URANS prediction exhibits a more organized radial particle distribution, with particles preferentially displaced toward the peripheral region of the section and reduced particle occupancy in the airway core. By contrast, the corresponding SBES distribution extends over a broader portion of the cross section and shows less persistent peripheral concentration, indicating stronger radial redistribution toward the core region. A similar contrast result is observed at CS1 near the bifurcation region. The URANS particle field contains relatively coherent particle-rich bands aligned with preferred transport pathways imposed by the local secondary-flow structure, whereas the SBES result exhibits weaker banding and a broader cross-sectional spread. These differences are consistent with the distinct treatments of turbulent particle transport in the two approaches. At CS3, where both models predict a pronounced high-velocity core, the particle distributions are comparatively similar, indicating that the influence of the resolved turbulent fluctuations on the instantaneous radial distribution is less pronounced at this location. Overall, SBES predicts weaker persistence of organized particle bands and a broader cross-sectional occupancy than URANS, as depicted in Fig.9.

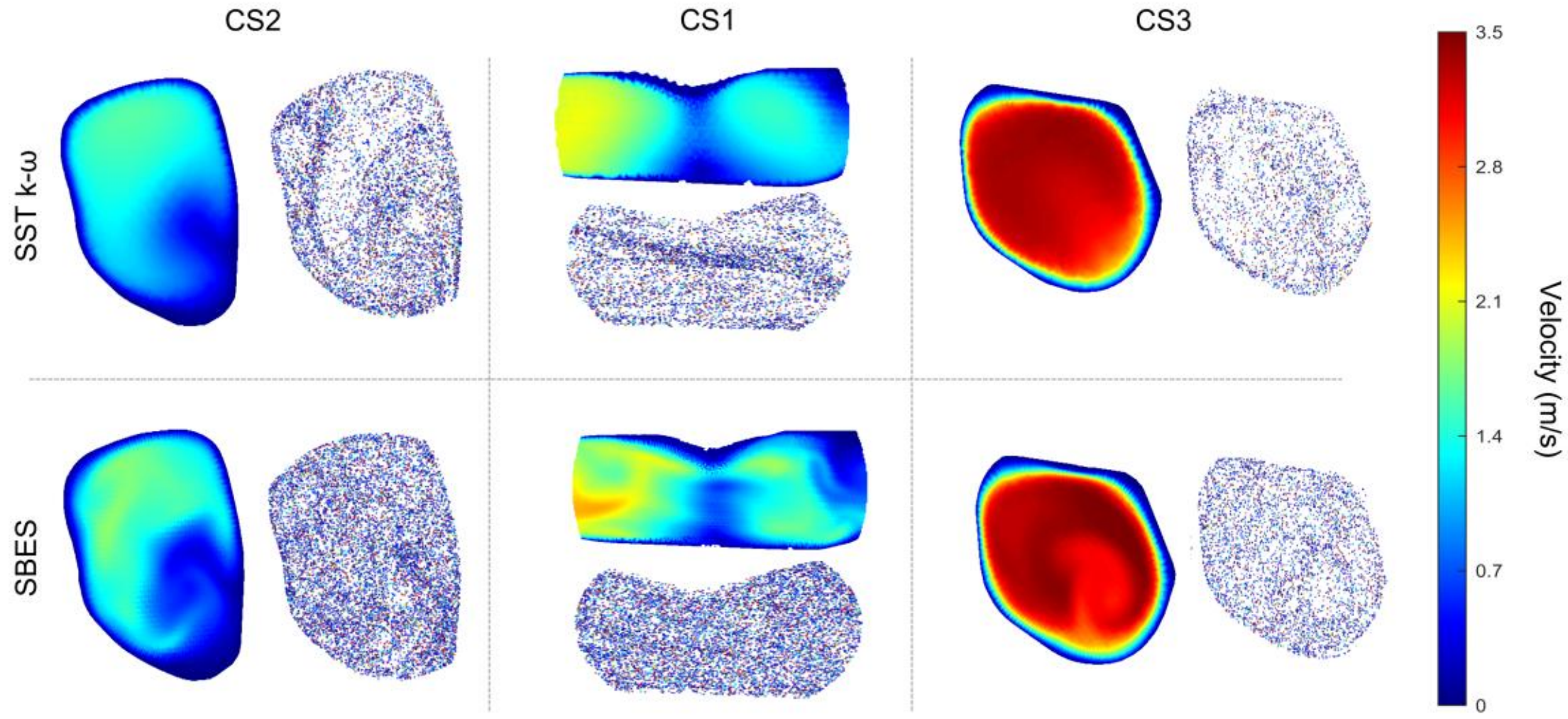

FIG. 9. Instantaneous velocity-magnitude fields and particle distributions on transverse sections CS2, CS1, and CS3, arranged from left to right. The upper and lower rows show the URANS and SBES results, respectively.

## D. Multi-scale quantification of wall deposition

Fig.10 presents an SBES-referenced spatial prediction-bias map of particle deposition on the airway walls across five representative anatomical regions, extending from the nasal cavity to the tracheobronchial tree. Blue regions denote deposition areas captured by SBES but omitted by URANS, orange regions denote additional areas predicted only by URANS, and gray regions represent the spatial overlap between the two predictions. The bias distribution exhibits a pronounced asymmetry. The omitted regions generally form spatially compact patches, indicating that URANS fails to reproduce portions of the localized deposition structures and high-concentration hotspots resolved by SBES. By contrast, the spurious URANS predictions are more broadly distributed and frequently adjoin or interweave with the matched regions. This spatial arrangement indicates that URANS not only misses localized deposition patches but also extends the predicted deposition footprint into neighboring wall regions. The coexistence of compact omissions and peripheral spurious predictions is consistent with spatial smearing induced by the Reynolds-averaged treatment of turbulent fluctuations, which broadens localized deposition patterns and reduces the spatial contrast of deposition hotspots.

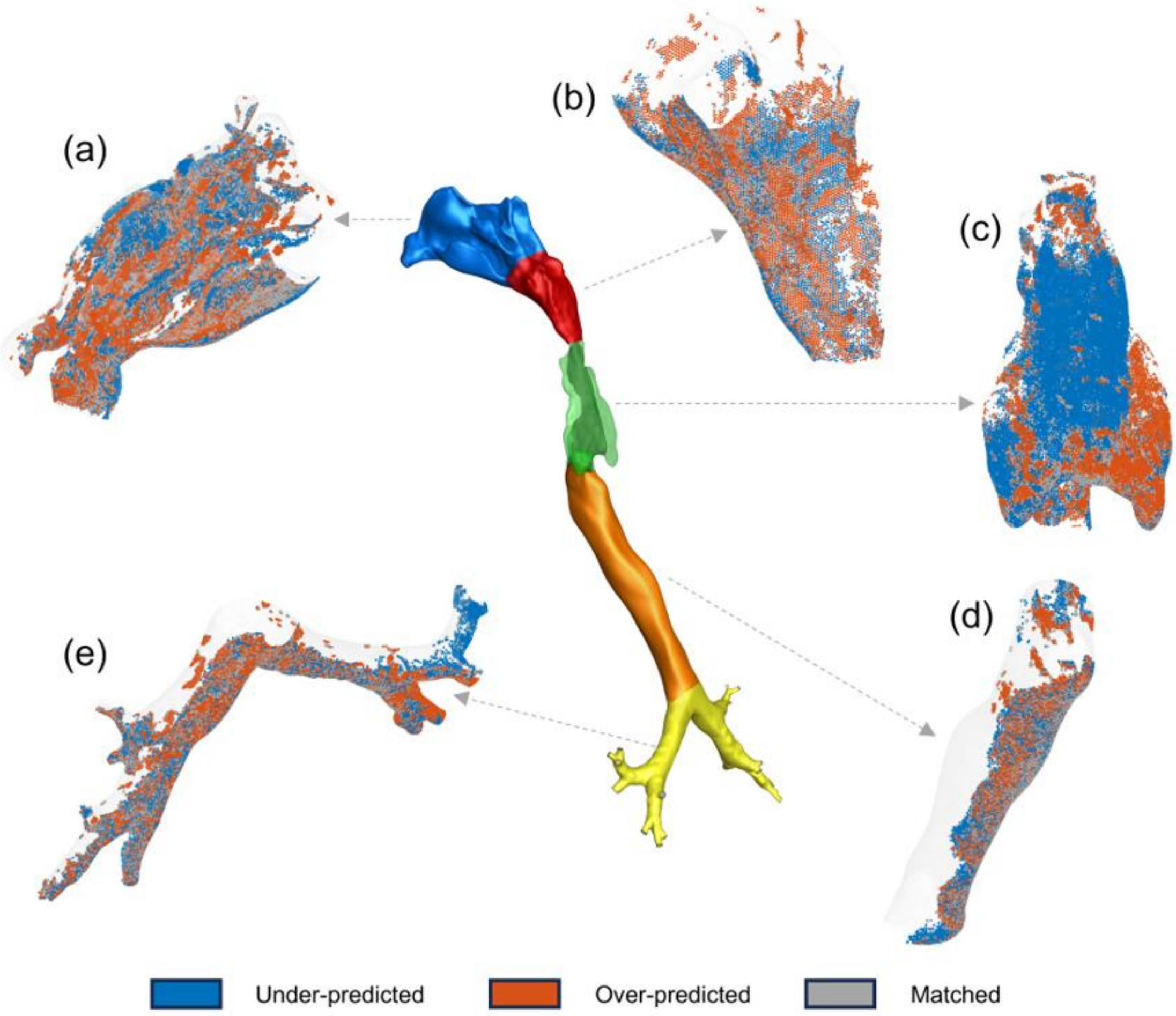

FIG. 10. SBES-referenced spatial classification of particle deposition on the airway walls across five representative anatomical regions. Panels (a–e) show the nasal cavity, nasopharynx, laryngopharynx, trachea, and tracheobronchial tree, respectively. Blue regions denote deposition captured by SBES but omitted by URANS, orange regions denote deposition predicted only by URANS, and gray regions denote spatially matched deposition.

Fig.11 quantifies the SBES-referenced prediction biases of URANS in terms of deposition area coverage and mass loading. The overall effective-deposition results are presented in Fig. 11a and Fig. 11b, whereas the corresponding results for the high-concentration hotspot domain are presented in Fig. 11c and Fig. 11d. Because the missed and false fractions are normalized by the SBES and URANS totals respectively, the two classes of metrics should be interpreted as complementary measures of omission and spurious prediction rather than as directly additive errors. At the overall-deposition level, the missed-area fraction ranges from 11% to 33%, with a regional mean of approximately 18%, while the missed-mass fraction ranges from 6% to 34%, with a mean of approximately 12%. The largest omission in deposition area occurs in the laryngopharynx, whereas the largest omitted mass fraction occurs in the bronchial region. The corresponding false-area and false-mass fractions range from 17% to 42% and from 5% to 29%, with regional means of approximately 29% and 20%, respectively. Thus, URANS reproduces a substantial portion of the overall deposition footprint but also predicts a relatively broad additional deposition area that is not supported by the SBES reference. Together with the spatial maps in Fig. 10, these results indicate spatial broadening of localized deposition patterns, consistent with the Reynolds-averaged treatment of turbulent fluctuations in URANS.

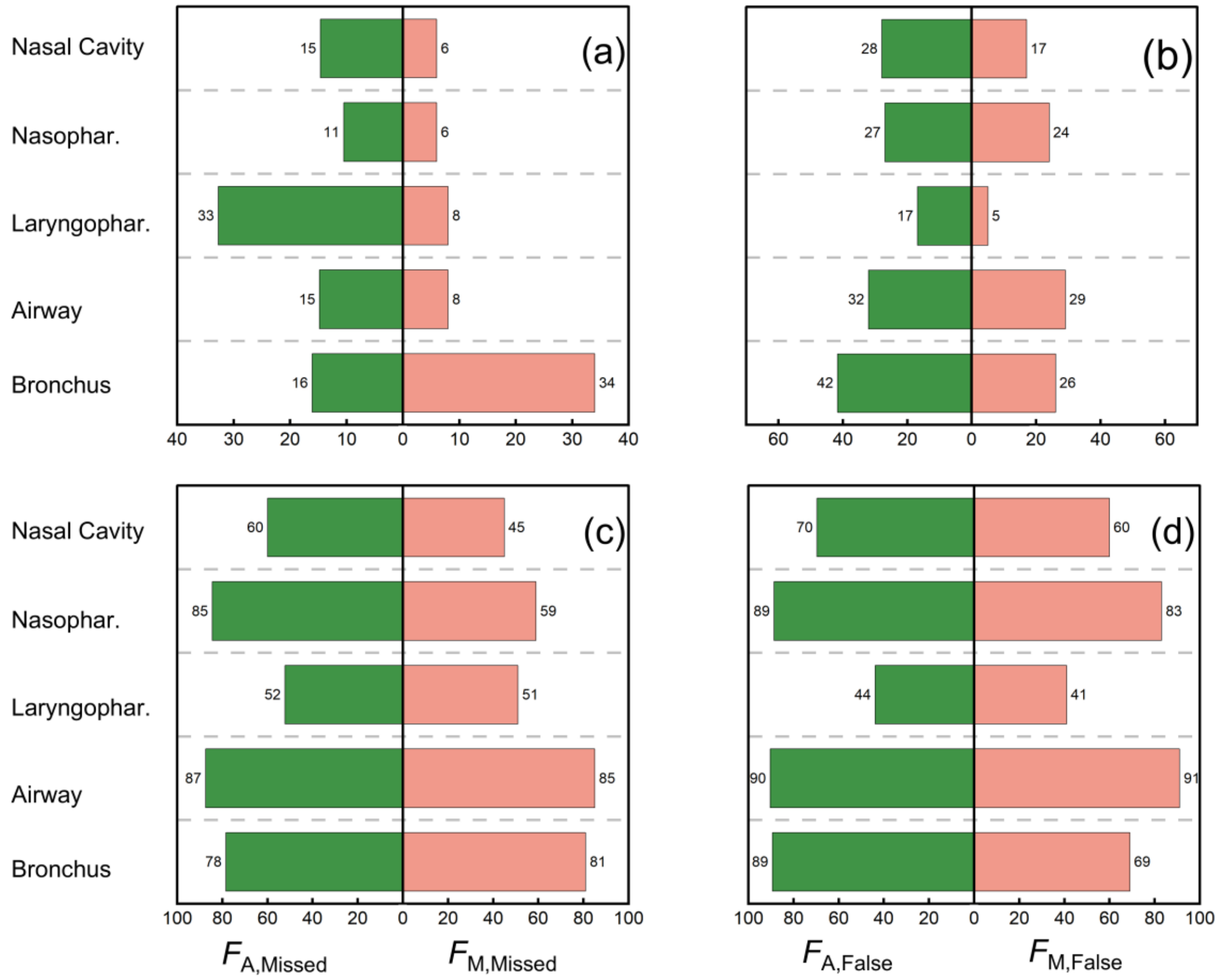

FIG. 11. Multi-scale spatial error quantification of particle deposition leveraging a dual-dimensional (geometric area and mass loading) framework. (a) Global-scale omissions (false negatives). (b) Global-scale spurious predictions (false positives). (c) Extreme hotspot omissions. (d) Extreme hotspot spurious predictions.

It is also depicted in Fig.11 that, the discrepancies become substantially larger within the high-concentration hotspot domain. The missed-area fractions range from 52% to 87%, and the missed-mass fractions range from 45% to 85%, with corresponding regional means of approximately 72% and 64%. Meanwhile, the false-area and false-mass fractions range from 44% to 90% and from 41% to 91%, averaging approximately 76% and 69%, respectively. These results show that URANS omits a large proportion of the hotspot area and deposited mass identified by SBES, while much of the hotspot footprint predicted by URANS does not spatially coincide with the SBES simulation. The combined area- and mass-based metrics therefore reveal a dual prediction bias: the overall deposition field is spatially broadened, whereas localized high-concentration hotspots are poorly reproduced in both position and mass loading. This behavior supports the conclusion that the Reynolds-averaged and isotropic eddy-viscosity treatment in URANS cannot reproduce the localized particle transport generated by the transient vortices and turbulent eddies resolved in SBES.

# IV. DISCUSSION

By establishing URANS and SBES models of the human respiratory tract, this study investigated the effects of transient vortices and turbulent eddies on the transport and deposition of drug particles within the respiratory tract during physiological-realistic breathing process. This study analyzed turbulent velocity fluctuations and turbulent energy spectra from the SBES results at typical locations in the respiratory tract during normal breathing, confirming the presence of turbulence conforming to Kolmogorov's -5/3 law, a high-resolution feature that URANS simulations fail to capture.[41, 43] Furthermore, by utilizing Voronoï diagrams, the spatial topological morphologies of drug particles at the laryngeal jet were analyzed, revealing significant topological differences between the SBES and URANS results. The SBES results confirmed the dominant role of transient vortices in driving the "preferential concentration" of particles. Subsequently, this study investigated the impact of complex vortex structures on the deposition patterns of particles on the respiratory tract walls through the visualization and quantitative analysis. This study provides useful support for accurately evaluating drug delivery efficiency of inhalation therapy and the risk of tissue complications.

This study conducted an in-depth comparison of the differences in the spatial distribution patterns of drug particles between the URANS and SBES models. The results indicate that the discrepancies between the two models originate primarily from their distinct capabilities in resolving small-scale turbulent eddies. Previous studies had investigated the coupling mechanisms between turbulent eddies and particle dynamics in channel and pipe flows[23, 26, 28, 44]. Marchioli et al. utilized direct numerical simulation (DNS) methods to study coherent structures near the wall, revealing the interaction between near-wall vortices and particles[32]. Eaton et al. studied the preferential concentration phenomenon under various flow conditions, finding that this preferential concentration phenomenon is significant in many types of flows and substantially affects the particle deposition process[29]. The findings about drug particles' transport and deposition in the respiratory tract in this study are consistent with the conclusions mentioned above: discrete micro-particles are centrifuged out of turbulent vortex cores and preferentially accumulate in high-strain regions, thereby producing a significant "preferential concentration" effect. This study introduced Voronoï topological analysis tools to quantitatively characterize the spatial topological features of this effect in human airways[28]. The results demonstrate that the SBES simulation reproduced the "preferential concentration" of particles in the respiratory tract, exhibiting complex cluster morphologies with high fractal dimensions ($D > 1.2$). These results confirmed the decisive influence of transient flow structures on the spatial distribution of particles. Meanwhile, the high-frequency turbulent fluctuations captured by the SBES model break the spatial confinement imposed by time-averaged secondary flows. This process transports particles toward the central void zones of sequential bifurcations and enables particles to disperse across the full cross-section of the bronchial passages.

With wall deposition visualization of particles and a set of newly-defined spatial accuracy metrics, this study further demonstrates the prediction biases induced by the time-averaging treatment of flow fields within the URANS approach. Prior investigations have identified bronchial bifurcations as high-risk regions prone to intensive particle deposition, a phenomenon that is potentially correlated with the onset of lung cancer.[10] While URANS simulations offer computational efficiency in predicting average respiratory deposition rates, the results of spatial accuracy metrics compared with SBES simulations demonstrated their failure at the local scale (Figure 11). Constrained by the excessive artificial turbulent diffusion introduced via the isotropic eddy viscosity assumption, the URANS model excessively smears particle distribution within the near-wall region. Without driving forces from physically resolved vortex structures, the predicted deposition area coverage and mass loading from URANS simulations deviate by approximately 20% relative to SBES results. More critically, URANS fails to capture the true high-risk hotspots corresponding to the top 5% particle concentration, and instead produces spurious particle accumulation along the airway walls.

The prediction deviations of URANS simulations regarding particle deposition distributions and local concentration extrema may impact the safety assessment of inhaled drugs[45-48]. Pharmacokinetic and inhalation toxicology studies indicate that efficacy bottlenecks and associated toxic side effects of inhalation therapies are often directly attributed to insufficient drug deposition in targeted lesions and extreme particle accumulation in non-targeted airway regions[10]. If URANS frameworks are adopted to guide the formulation of clinical aerosol dosing regimens, their time-averaged particle deposition predictions tend to overestimate the effective spatial coverage of targeted pharmaceutical agents while underestimating localized tissue toxicity within high-deposition hotspots. This study confirmed that employing high-fidelity scale-resolving SBES simulations better reveals spatial particle distributions, wall deposition morphologies, and high-concentration deposition regions, thereby providing a methodological support for optimizing targeted inhalation delivery protocols and evaluating local tissue dosage safety.

Although the comparative analysis of the SBES and URANS simulations provides novel insights, certain physical simplifications remain within the current model. First, the computational domain assumes rigid airway boundaries, neglecting airway compliance and the associated fluid-structure interaction (FSI) effects[49-53]. Moreover, ciliary kinematics and wall roughness investigated by Cui et al. and Kong et al. were not included [54, 55]. Second, the drug particles were idealized as spherical and omitted the influence of complex particle morphology on deposition dynamics. [56, 57] Third, the present study restricts the inhalation route to the nasal cavity, precluding the specific oropharyngeal aerodynamic characteristics unique to oral delivery devices such as dry powder inhalers (DPIs) (Inthavong et al.)[58]. Future investigations will integrate these multiphysics effects alongside high-fidelity *in vitro* experimental validation.

# V. CONCLUSION

In summary, this study elucidates the fundamental physics of vortex-driven aerosol dynamics within an anatomically realistic human airway. By juxtaposing high-fidelity SBES simulation against URANS simulations, it was demonstrated that explicitly resolving transient, cross-scale turbulent eddies is a critical prerequisite for accurately predicting localized deposition hotspots. The Voronoï topological analysis proves that high-frequency turbulent fluctuations induce intense radial dispersion. This dynamic back-mixing disrupts the confinement of mean secondary flows, enabling particles to penetrate the central void regions across anatomical strictures (characterized by cluster fractal dimensions of $D > 1.2$). Conversely, the time-averaging kernel inherent to URANS artificially smears these transient structures, yielding a state of spurious uniformity. Leveraging the proposed asymmetric error framework, we quantify that this physical simplification introduces a critical dual deficiency: it globally over-smears near-wall deposition footprints by approximately 20%, while simultaneously failing to capture genuine extreme hotspots (the top 5% concentration extrema) and generating non-physical topological artifacts.

# ACKNOWLEDGMENTS

This study was supported by the National Key Research and Development Program of China (2024YFC2419005).

# AUTHOR DECLARATIONS

## Conflict of Interest

The authors have no conflicts to disclose.

## Author Contributions

**Mengtao Li :** review and editing (lead); conceptualization (lead); data curation (lead); formal analysis (lead); investigation(lead); methodology(lead); validation(lead); visualization(lead); writing – original draft (lead); software (lead). **Yawei Wang:** project administration(lead); resources(lead); supervision (lead); funding acquisition(lead)writing – original draft (supporting); writing – review and editing (supporting); methodology(equal); Funding. **Wentao Feng:** supervision(supporting). **Yubo Fan:** supervision(supporting).

# DATA AVAILABILITY

The data that support the findings of this study are available from the corresponding author upon reasonable request.